\documentstyle[sprocl]{article} 
 \input{psfig} \arraycolsep1.5pt 
\begin{document}
\title{ON THE EFFECTIVE LIGHT-CONE QCD-HAMILTONIAN}
\author{H.C. PAULI}
\address{Max-Planck-Institut f\"ur Kernphysik, 
         D-69029 Heidelberg, Germany. \\E-mail:pauli@mpi-hd.mpg.de}
\maketitle\abstracts{ 
  Taking the effective interaction between a quark and an anti-quark
  from previous work, the dependendence on a regularization scale 
  is removed in line with the renormalization group.
  In order to emphasize the essential point, the full spinor 
  interaction is replaced by a model which includes only the 
  Coulomb and the hyperfine interaction. By adjusting the effective
  quark masses, the only free parameters of the theory, 
  the mass and the size of the pion are reproduced, 
  as well as the mass of all other pseudo-scalar mesons.
  Estimates for the vector mesons are close to the empirical values.
  The model exposes screening rather than strict confinement.
  The ionization thresholds are in general much larger than the pion mass.
}\hyphenation{ana-lyz-ed any-more ap-pro-xi-ma-tion can-not
   chro-mo-dy-na-mics 
   ei-gen-func-tion ei-gen-val-ue Ha-mil-to-ni-an 
   re-cog-ni-ze re-la-ti-vis-tic re-nor-ma-li-za-ti-on theo-rems
}
\section{Introduction}

The bound-state problem in a field theory particularly 
gauge field theory is a very difficult many-body problem 
and has not been solved thus far.
But perhaps an exact solution is not necessary
for reconciling the so succesful constituent quark models
with a covariant theory such as QCD and for being
able to compute observables like the structure functions
and distribution amplitudes from an underlying theory.
Perhaps it suffices to have approximate or 
QCD-inspired solutions which are not wrong from the outset.
The present work is still ongoing and contributes to these questions.
Perhaps one begins to understand the essential point.

\section{A QCD-inspired effective interaction}

The light-cone approach to the bound state problem aims at 
diagonalizing the invariant mass-squared operator.\cite{BroPauPin98}
Its matrix elements are obtained directly 
from the gauge field Lagrangian in the light-cone 
gauge and are explicitly tabulated.\cite{BroPauPin98} 
The many-body aspect of the problem was reduced recently, 
by the method of iterated resolvents,\cite{Pau98}  
to the problem of phrasing an effective Hamiltonian 
which acts only in the Fock space of one quark and one anti-quark. 
The reduction is not exact but rests on certain simplifying 
assumptions which can be checked only {\it a posteriori}.
Quite cautiously, one should therefore speak of 
a `QCD-inspired' effective Hamiltonian. 
Important is that both the effective and the full Hamiltonian
have in principle the same eigenvalue spectrum and that
the Fock space is systematically reduced to the $q\bar q$-space.
Important is also that the Fock-space is not truncated, 
that all Lagrangian symmetries are preserved,
and that the higher Fock-space amplitudes can be retrieved
systematically from the $q\bar q$-projection once that is available.

\begin{figure} [t] 
\begin{minipage} [b] {55mm}
  \psfig{figure=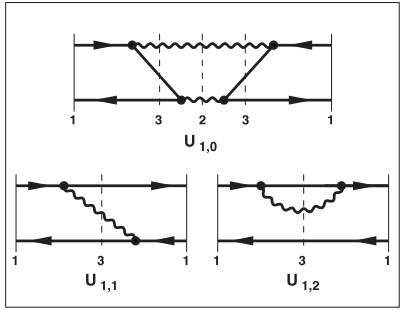,height=1.6in}
\end{minipage}
\hfill
\begin{minipage} [b] {60mm}
\caption{\label{fig:2.1}
  The three graphs of the effective interaction in the 
  $q\bar q$-space according to \protect\cite{Pau98}.
  The upper graph illustrates an effective interaction due to a 
  two gluon annihilation. For flavor-off-diagonal mesons
  its amplitude is kinematically suppressed. The lower 
  two graphs correspond to an effective one gluon exchange.~--
  The effective interaction scatters a quark with four momentum 
  and helicity $(k_{1},\lambda_{1})$ into $(k'_{1},\lambda'_{1})$,
  and correspondingly the anti-quark from $(k_{2},\lambda_{2})$ 
  into $(k'_{2},\lambda'_{2})$. 
}\end{minipage}
\end{figure}

The effective interaction has several contributions
which are illustrated in Fig.~\ref{fig:2.1}.
The present work is restricted to flavor-off-diagonal mesons,
to mesons where quark and anti-quark have different flavors.
Then the effective interaction due to the annihilation
of two gluons is kinematically suppressed, 
and the final (one-body) integral equation in the $q\bar q$-space 
is governed by an effective one gluon exchange, {\it i.e.}
\begin{eqnarray} 
    M^2\langle x,\vec k_{\!\perp}; \lambda_{1},
    \lambda_{2}  \vert \psi\rangle = 
    \left[ 
    \frac{\overline m^2_{1} + \vec k_{\!\perp}^{\,2}}{x} +
    \frac{\overline m^2_{2} + \vec k_{\!\perp}^{\,2}}{1-x}  
    \right]\langle x,\vec k_{\!\perp}; \lambda_{1},
    \lambda_{2}  \vert \psi\rangle\phantom{.}&& 
\nonumber\\ 
    - {1\over 4\pi^2}
    \sum _{ \lambda_q^\prime,\lambda_{2}^\prime} \!\int 
    \frac{ dx^\prime d^2 \vec k_{\!\perp}^\prime }
    {\sqrt{ x(1-x) x^\prime(1-x^\prime)}}
    R(x',\vec k'_{\!\perp};\Lambda)
    \times 
    \phantom{\langle x^\prime,\vec k_{\!\perp}^\prime; 
    \lambda_{1}^\prime,\lambda_{2}^\prime  
    \vert \psi\rangle.}&&
\nonumber\\ 
    \times 
    \frac{4}{3}\frac{\overline\alpha(Q;\Lambda)}{Q^2} 
    \langle
    \lambda_{1},\lambda_{2}\vert S\vert 
    \lambda_{1}^\prime,\lambda_{2}^\prime\rangle
    \,\langle x^\prime,\vec k_{\!\perp}^\prime; 
    \lambda_{1}^\prime,\lambda_{2}^\prime  
    \vert \psi\rangle.&&
\label{eq:1}\end{eqnarray} 
Here, $M ^2$ is the invariant-mass squared eigenvalue and
$\langle x,\vec k_{\!\perp};\lambda_{1},\lambda_{2}\vert\psi\rangle$ 
the associated eigenfunction. It is the probability amplitude for
finding a $q\bar q$-Fock state in which the quark has longitudinal 
momentum fraction $x$ and transversal momentum $\vec k_{\!\perp}$   
and the anti-quark correspondingly $1-x$ and $-\vec k_{\!\perp}$, and 
where the respective helicities are $\lambda_{1}$ and $\lambda_{2}$.
The mean four-momentum transfers of the quarks 
and the spinor factor are respectively defined by
\begin{eqnarray}
   Q ^2 &=& -\frac{1}{2}\left[(k_{1}-k_{1}')^2 +
            (k_{2}-k_{2}')^2\right]
,\\ 
   S = \langle\lambda_{1},\lambda_{2}\vert S\vert 
   \lambda_{1}^\prime,\lambda_{2}^\prime\rangle 
   &=&
   \left[ \overline u (k_{1},\lambda_{1})\gamma^\mu
   u(k_{1}^\prime,\lambda_{1}^\prime)\right] \, 
   \left[ \overline v(k_{2}^\prime,\lambda_{2}^\prime) \gamma_\mu 
    v(k_{2},\lambda_{2})\right] 
.\end{eqnarray}
The effective quark masses $\overline m _{1,2}$ and 
the effective coupling constant $\overline\alpha(Q;\Lambda)$ 
depend both, in general, on a regularization scale $\Lambda$.
The regulator function $R(x',\vec k'_{\!\perp};\Lambda)$ 
restricts the range of integration and in general is
a function of the same $\Lambda$, see also below.

The current work is restricted to the lowest order of approximation (LOA) 
where $\overline m  = m$ and $\overline \alpha(Q;\Lambda) = \alpha$.
The expressions for the next-to-lowest order (NLO) can be found
elsewhere.\cite{Pau98}
One has a fair confidence into the validity of Eq.(\ref{eq:1}),
since the alternative method of Hamiltonian flow \cite{Weg94} 
leads to the very same equation as in LOA.\cite{GubPauWeg98}  
The same equation had also been obtained prior to 
that,\cite{KPW92,TriPau97} with however less stringent arguments.
The work of Wilson {\it et al.} \cite{WilWalEta94}
had the same emphasis, but that it did not lead to the same 
formulas, see also Refs.\cite{BriPer96,BriPerWil97}.

\section{The crucial point in a simple model}

Why is a regularization necessary at all? 

In light-cone parametrization, the quark is at rest
relative to the anti-quark when $\vec k _{\!\perp}= 0$ and 
$ x = \overline x \equiv \overline m_1/(\overline m_1+\overline m_2)$.
For very small deviations from these `equilibrium values' 
the spinor factor is diagonal in the helicities, 
$\langle\lambda_{1},\lambda_{2}\vert S\vert 
 \lambda_{1}^\prime,\lambda_{2}^\prime\rangle 
 \sim 4\overline m_1\overline m_2
 \,\delta_{\lambda_1,\lambda_1'}\,\delta_{\lambda_2,\lambda_2'} $, 
as can be veryfied from an
explicit presentation of $S$.\cite{TriPau97}    
For very large deviations, particularly for
$\vec k_{\!\perp}^{\prime\,2} \gg \vec k_{\!\perp} ^{\,2}$,
holds  $Q ^2 \simeq\vec k_{\!\perp}^{\prime\,2}$ and 
$\langle\uparrow\downarrow\vert S\vert\uparrow\downarrow\rangle \simeq
 2\vec k_{\!\perp}^{\prime\,2}$.
The ratio is then a simple dimensionless number, $S/Q ^2 = 2$, 
independent of $\vec k'_{\!\perp}$.
Both extremes are combined here by
\begin{eqnarray}
   \frac{\langle\uparrow\downarrow\vert S\vert\uparrow\downarrow\rangle}
   Q ^2 = \frac{4\overline m_1\overline m_2}{Q ^2} + 2 
,\label{eq:4}\end{eqnarray}
which is considered as a {\em model of the interaction} in singlet
states, dropping the less important spin-orbit interaction.

Unfortunately, I have not realized earlier \cite{KPW92,TriPau97}
that the innocent and finite number `2' in Eq.(\ref{eq:4})
has the same orign as the familiar divergencies in gauge field
theory. The latter arise, typically, when bilinar products of the
elementatry Dirac interaction like in $S$ divided by
some energy denominator like $Q ^2$ are integrated over all
phase space. 

It was clear from the outset that the many-body Hamiltonian 
has to be regularized.\cite{KPW92} 
Since fancy methods like dimensional regularization are
not applicable, Fock-space regularization
was considered to be sufficient.\cite{BroPauPin98} 
In practice, it has lead to difficulties and was
was replaced by vertex regularization,\cite{Pau98} 
where the elementary Dirac interaction at a vertex was endorsed 
with kind of a form factor $F(\Lambda)$, {\it i.e.}
\begin{eqnarray}
    \left[\overline u(1)\gamma^\mu u(2)\right] \epsilon_\mu(3)
    \longrightarrow 
    \left[\overline u(1)\gamma^\mu u(2)\right] \epsilon_\mu(3)
    \,F(k_1,k_2,k_3;\Lambda)
.\end{eqnarray}
Requiring that the off-shell mass of the two scattered particles
$M_0^2 = (k_2+k_3)^2$ is limited, 
$F$ becomes the step function
$F(k_1,k_2,k_3;\Lambda) = \Theta(M_0^2 - \Lambda^2)$.
In the absolute squares of the effective interaction, 
as in Eq.(\ref{eq:1}), it appears as a regulator function R
as a consequence of regulating the theory from the outset.

\section{Rewriting the integral equation in conventional momenta}

Unpleasant is that $x$ and $\vec k_{\!\perp}$ have different ranges 
$(0 \leq x \leq 1,\ -\infty \leq k_{x} \leq \infty)$. 
Instead of $x$  a $k_z$ 
$(\ -\infty \leq k_{z} \leq \infty)$ can be introduced
by the transform \cite{Saw85}
\begin{eqnarray}
    x(k_z) &=& \frac{E_1+k_z}{E_1+E_2},\qquad\mbox{with\ }
    E_{1,2} = \sqrt{\overline m^{\,2} _{1,2}+\vec{k}^{\,2} _{\!\perp}+k^2_z}
.\end{eqnarray}
The Jacobian of the transformation can be cast into the form
\begin{eqnarray}
    \frac {dx}{x(1-x)} &=& \frac {1} {A ({k})}\frac {dk_z}{m_r} 
    ,\qquad\mbox{with\ }
    A (k) = \frac{1}{m_r} \frac{E_1E_2}{E_1+E_2} 
,\\
   \mbox{and}\qquad
   \frac{1}{m_r} &=& \frac{1}{\overline m_1} + \frac{1}{\overline m_2} 
   ,\qquad
   m_s = \overline m_1 +\overline m_2
.\end{eqnarray}
In the same convention the kinetic energy becomes
\begin{eqnarray}
    T (k) &=&  
    \frac{\overline m^2_{1} + \vec k_{\!\perp}^{\,2}}{x} +
    \frac{\overline m^2_{2} + \vec k_{\!\perp}^{\,2}}{1-x} - m_s^2 
    \quad\equiv \quad C(k)\,\vec k ^2
,\\
    \mbox{with}  \quad
    C(k) &=& (E_{1}+\overline m_{1} + E_{2}+ \overline m_{2}) \left(
    \frac{1}{E_{1}+\overline m_{1}} +
    \frac{1}{E_{2}+\overline m_{2}}\right) 
,\end{eqnarray}
without approximation. If one substitutes the wave function according to 
\begin{eqnarray}
   \psi(x,\vec{k}_{\!\perp}) &=&
   \sqrt{\frac{A(x,\vec{k}_{\!\perp})}{x(1-x)}} 
   \ \phi (x,\vec{k}_{\!\perp})
\end{eqnarray}
and drops explicit reference to the helicities, one arrives at
an equation,
\begin{eqnarray}
   \left[M^2 - m_s^2 - C(k)\vec{k}^2 \right]\phi(\vec{k})=-
   \frac{1}{4\pi^2 m_r} \int
   \frac {d^3\vec{k'} \ R} {\sqrt{A(\vec k)A(\vec k ^\prime)}} 
   \ \frac{4}{3}\frac{\overline\alpha S}{Q^2} 
   \,\phi(\vec{k'})
,\label{eq:12}\end{eqnarray}
which is identical with Eq.(\ref{eq:1}), 
but which looks like one with usual 3-momenta. 

\section{Interpretation in configuration space}

The main reason for introducing conventional 3-momenta
is interpretation. 
It is more transparent in configuration space.
Since the various factors $A(\vec k)$ with their square-roots
prevent straight-forward Fourier transforms,
the non-relativistic approximation 
($\vec k ^{\,\prime\,2} \ll \overline m ^2_{1,2}$)
is applied consistently.
Using
\begin{eqnarray}
   A (k) = 1
,\qquad
   C (k) = \frac{m_s}{m_r}
,\qquad
   Q^2   = (\vec k - \vec k')^2
,\end{eqnarray}
the model interaction of Eq.(\ref{eq:4}), 
and in addition the regulator $R=1$, 
Eq.(\ref{eq:12}) leads to a local Schr\"odinger equation
\begin{eqnarray}
   \left[M^2-m_s^2-\frac{m_s}{m_r}\vec{k}^2 \right]\phi(\vec{k}) &=& - 
   \frac{4\alpha}{3\pi^2}\int d^3\vec{k'} 
   \left(\frac{m_s}{Q^2} + \frac{1}{2m_r}\right) 
   \,\phi(\vec{k'})
,\\
   \left[M^2-m_s^2-\frac{m_s}{m_r}\vec p^{\,2}\right]\psi(\vec r) &=& 
   2m_s V(r) \psi(\vec r)
   ,\quad\mbox{ with } \vec p \equiv i\vec\nabla
.\end{eqnarray}
The potential $V(r)$ bears great similarity with $V_{\mbox{hf-s}}(r)$,
the hyperfine interaction in the singlet channel found for hydrogen
in all textbooks, {\it i.e.}
\begin{eqnarray}
   V (r) &=&   - \frac{4}{3}\alpha\left(\frac{1}{r} + 2\phantom{_p}
   \frac{\pi }{m_r m_s}\delta^{(3)}(\vec r)\right)
,\label{eq:16}\\
   V_{\mbox{hf-s}} (r) &=&  - \phantom{\frac{4}{3}}\alpha\left(\frac{1}{r} +  
   g_p\frac{\pi }{m_e m_p}\delta^{(3)}(\vec r)\right)
.\end{eqnarray}
The strange `2' in Eq.(\ref{eq:4}) finds its explanation 
as the gyromagnetic ratio for a fermion, with $g_p=2$.
But Eq.(\ref{eq:16}) has no solution! 
A Dirac-delta function is no proper function and must be regulated,
for instance by a Yukawa-potential 
\begin{eqnarray}
   V (r) &=& - \frac{4\alpha}{3} \left( \frac{1}{r} +
   \frac{\mu^2}{m_r m_s}\, \frac{e^{-\mu r}}{r}\right)
.\end{eqnarray}
Important is that the delta ($\int d^3\vec r \,\delta^{(3)}(\vec r) = 1$) 
and the Yukawa have the same strength 
$\int d^3\vec r \,[\mu^2 \mbox{exp}(-\mu r)/(2\pi r)]=1$.
Transforming back to momentum space gives
\begin{eqnarray}
   \left[M^2-m_s^2-\frac{m_s}{m_r}\vec k^2\right]\phi(\vec{k}) = - 
   \frac{4\alpha}{3\pi^2}\int\!\! d^3\vec{k'} 
   \left(\frac{m_s}{Q^2} + \frac{1}{2m_r}
   \frac{\mu^2}{\mu^2+Q^2} \right)\phi(\vec{k'})
.\label{eq:20}\end{eqnarray}
Obviously, replacing in configuration space the Dirac-delta function 
with a Yukawa potential 
corresponds to introducing in momentum space the regulator function
$R = \mu^2/(\mu^2+Q^2)$.

The regularization of a Dirac-delta function is an old theme
of nuclear physics in the context of pairing theory.\cite{BraDamJen72} 
It was the point of orign for the 
similarity transform,\cite{GlaWil93,GlaWil94} 
and was investigated recently,\cite{PerSzp99} again.

\begin{figure}  
\begin{minipage} [t] {55mm}
  \psfig{figure=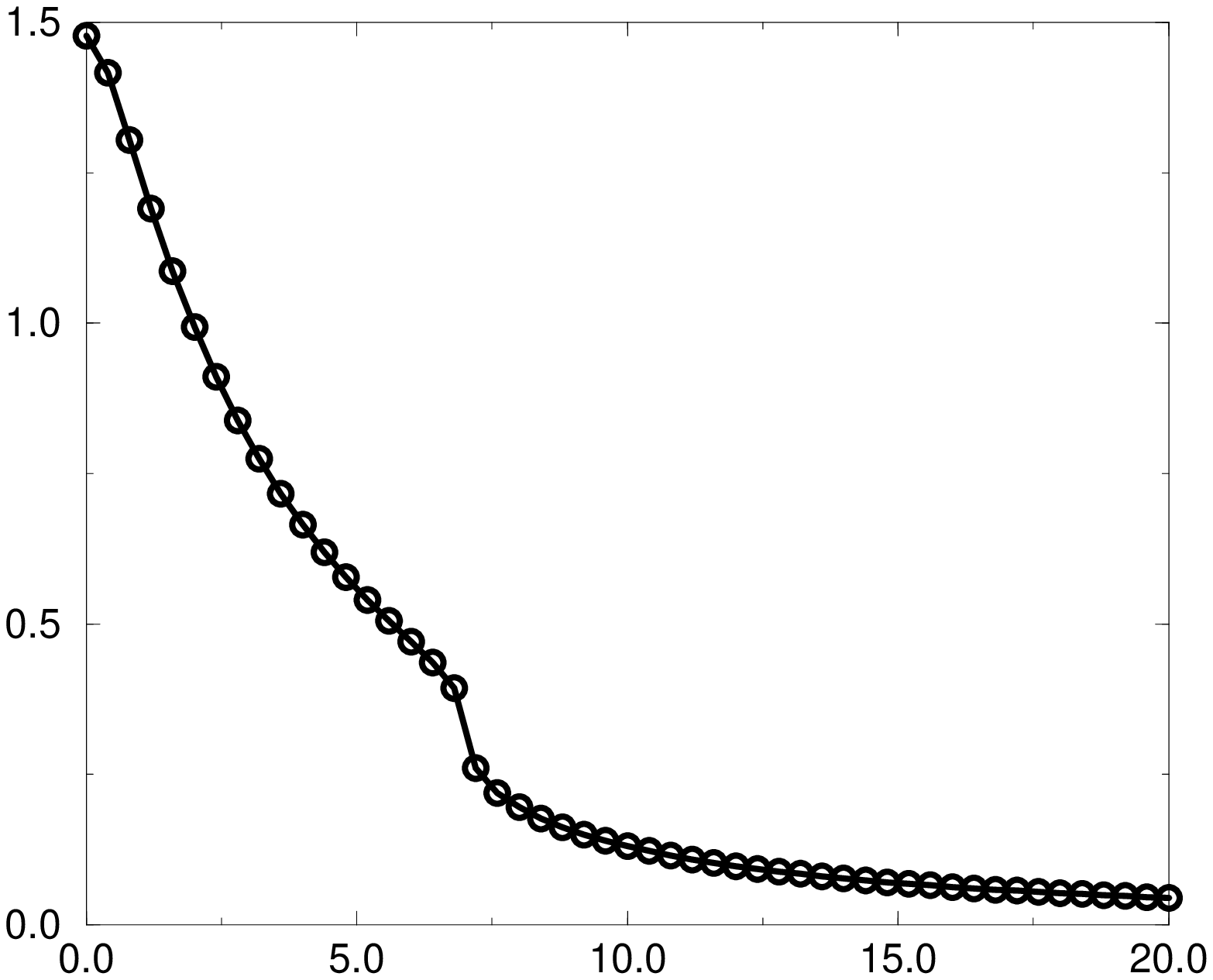,height=1.8in}
\caption{\label{fig:5.1}  
   Coupling constant $\alpha$ versus regularization scale $\mu$ 
   in units of 350 MeV.
}\end{minipage}
\hfill
\begin{minipage} [t] {55mm}
  \psfig{figure=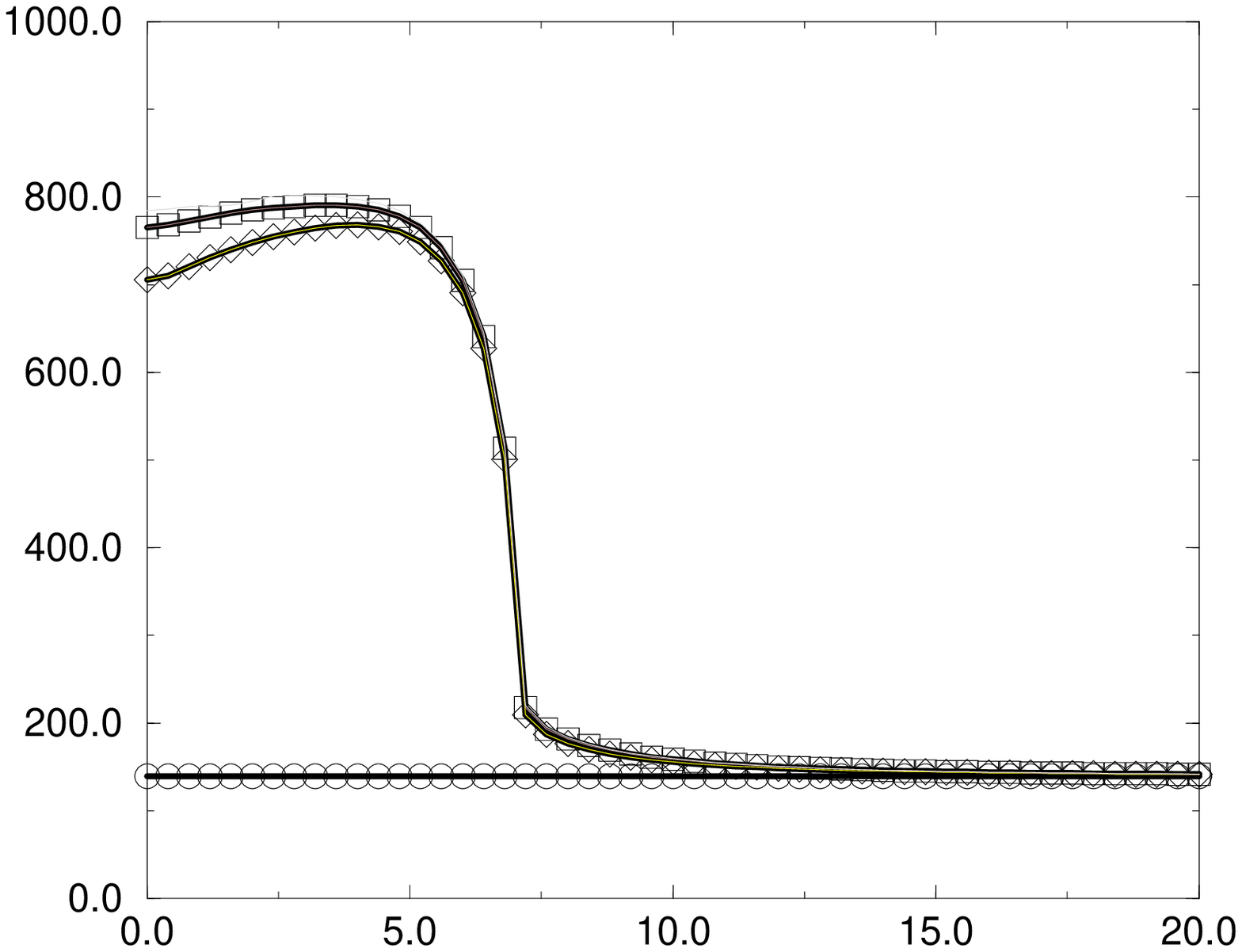,height=1.8in}
\caption{\label{fig:5.2}  
   The spectrum of the first three singlet-s states in MeV 
   versus regularization scale $\mu$ in units of 350 MeV.
}\end{minipage}
\end{figure}

\section{Renormalization}
\label{sec:renormalization}

The eigen values of the integral equation Eq.(\ref{eq:20}) depend 
now on a regularization scale $\mu$.
In line with the current understanding of field theory
one replaces $\alpha$ and $m$ by functions of $\mu$,
such that every eigenvalue is stationary against small
variations of $\alpha$, $m$ and $\mu$, {\it i.e.} $dM_n^2/d\mu = 0$!
The general procedure, however, is not possible here,
since $\alpha$ and $\mu$ appear in Eq.(\ref{eq:20}) in the
typical combination $\alpha\mu^2/(\mu^2+Q^2)$.
One therefore proceeds by looking for a function $\alpha_{\mu}$ such,
that the {\em calculated} mass of the pion, for example,
agrees with the empirical value.
The so obtained function $\alpha_{\mu}$ is considered universal.
The actual value of $\mu$ can then be fixed by a second requirement.

\section{The pion}

For carrying out, in practice, the programme of 
Sec.~\ref{sec:renormalization}
one needs an efficient tool for solving Eq.(\ref{eq:20}). 
Such one has been developed recently,\cite{BieIhmPau99} 
restricted however for simplicity to spherically symmetric s-states.
Fixing the up and the down mass to $\overline m_u = \overline m_d=1.16$
(the unit of mass is chosen here as 350 MeV$/c^2$), 
the calculated pion mass is required to agree with the experimental 
value \cite{PDG98} to within eight digits.
For every value of $\mu$ one obtains thus an $\alpha_\mu$ which
is displayed in Fig.~\ref{fig:5.1}.
The even greater surprise came when investigating the spectrum
of the pion along the so obtained $\alpha_\mu$,
as given in Fig.~\ref{fig:5.1}. 
The lowest state, the $\pi^+$, stays nailed fixed to the empirical value,
while the second and the third (as well as the higher ones)
have an extremum at $\mu\simeq 3.8$. Such an extremum was not really
expected, but in view of the renormalization group is useful:
$\mu$ determines itsself from the solution!

For suffiently large values of $\mu$, say for $\mu \gg 10$,
see Fig.~\ref{fig:5.2}, the coupling constant decreases strongly
and the spectrum becomes more and more the familiar Bohr spectrum 
with a small hyperfine shift. 
Obviously, it can be calculated perturbatively, indeed,
for sufficiently small $\alpha$. 

\begin{figure}  
\begin{minipage} [t] {55mm}
  \psfig{figure=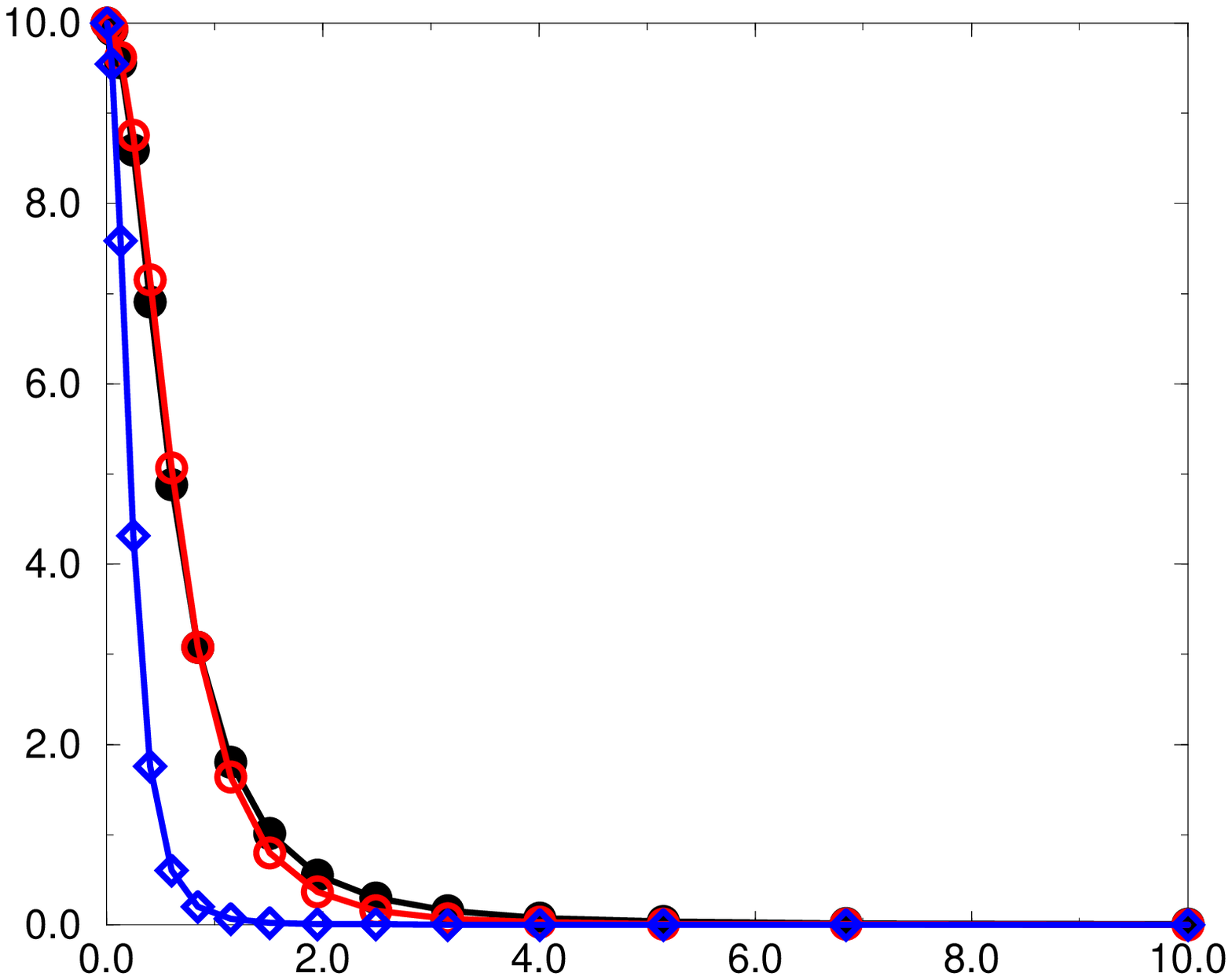,height=1.8in}
\caption{\label{fig:6.1}  
   The pion wave function $\Phi(p)$ 
   is plotted versus $p/1.552$ (filled circles)
   in an arbitrary normalization.
   The diamonds denote a Coulomb wave function with the same $\alpha$ 
   and the open circles an approximate wave function, see text.
}\end{minipage}
\hfill
\begin{minipage} [t] {55mm}
  \psfig{figure=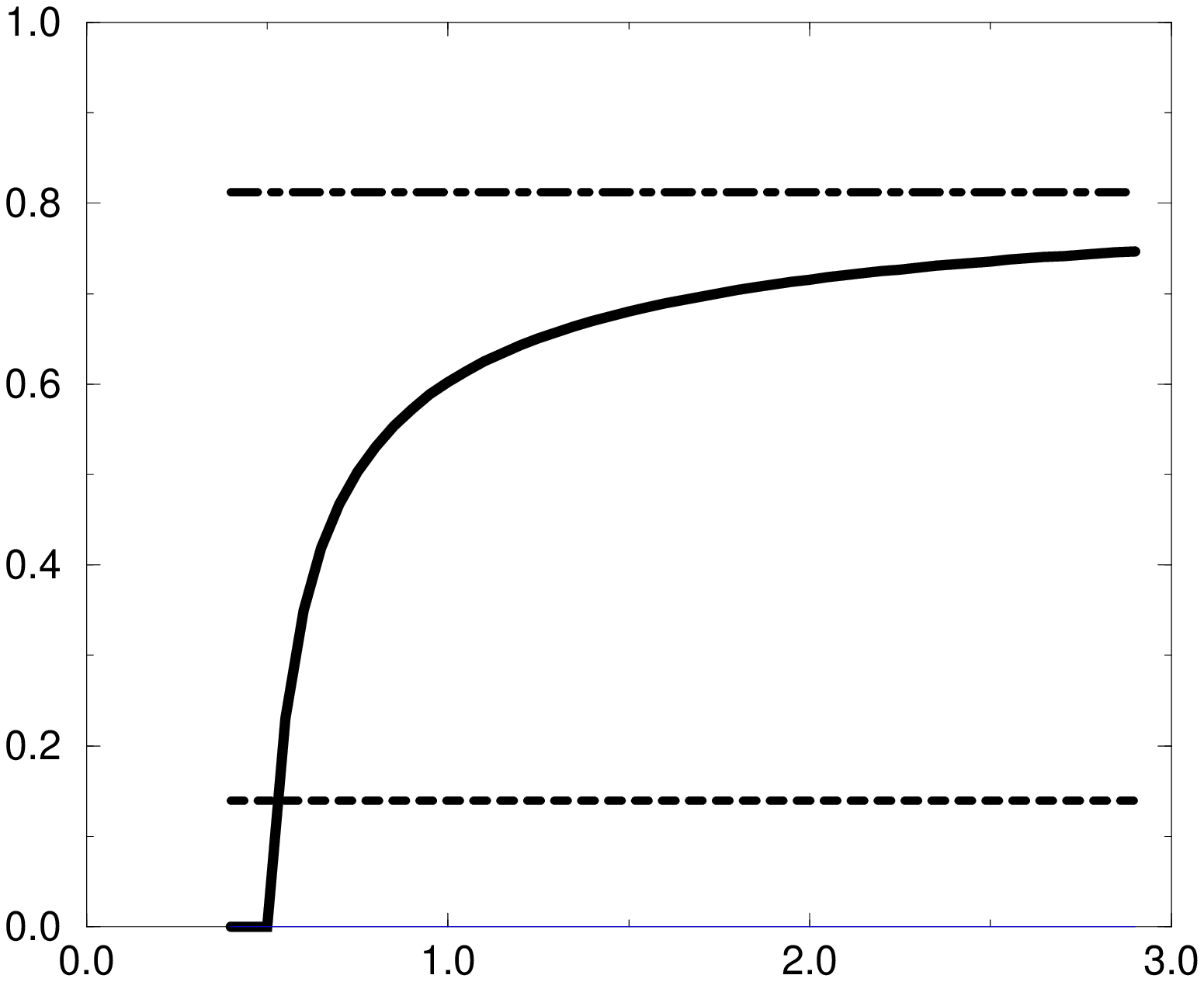,height=1.8in}
\caption{\label{fig:6.2} 
   The relativistic potential energy $W(r)$ for the pion is plotted 
   in GeV versus the radius in fm (solid line).
   The pion mass is indicated by the dashed, 
   and the ionization threshold by the dashed-dotted line.
}\end{minipage}
\end{figure}

Having fixed now the regularization scale $\mu\simeq\mu_0=3.8$
gives the coupling constant $\alpha\simeq\alpha_{\mu_0}= 0.6904$,
see also Fig.~\ref{fig:5.1}.
The wave function of the pion is plotted in Fig.~\ref{fig:6.1}.
By a fit to an approximate expression $\Phi_a(p) = 10/(1+p^2/p_a^2)$, 
a width $p_a=1.471$ is obtained which is much larger than 
$p_c=\frac{4}{3}\alpha m_r=0.5339$, the Bohr momentum of the Coulomb wave 
function with the same $\alpha$.
Having the wave function, one could, in principle,
calculate the form factor and the distribution amplitude
according to the exact expressions.\cite{BroPauPin98} 
This is however not a trivial undertaking, and, instead of,
the above approximate expression was Fourier transformed to 
$\psi(\vec r) \sim e^{-p_a r}$. The calculated root-mean-sqare radius
of the pion becomes then
\begin{equation}
    <r^2>^\frac{1}{2} = \frac{\sqrt{3}}{p_a} = 0.663 \mbox{\ fm}
.\label{eq:r}\end{equation}
The good agreement with the experimental value
$<r^2>^\frac{1}{2}_{\pi}=0.657$ fm\cite{PDG98} 
is actually the result of fitting the quark masses
$\overline m_u = \overline m_u = 406$ MeV, see above.
This exhausts all freedom in choosing the parameters of Eq.(\ref{eq:20}).

One should emphasize that the Yukawa potential in Eq.(\ref{eq:20})
at the scale $\mu_0$ pulls down essentially only one state.
The others are left rather unperturbed at their Bohr values,
see Fig.~\ref{fig:5.2}.
The analogy with the pairing model  
in early nuclear physics is more than obvious.\cite{BraDamJen72} 
In a future and more rigorous solution of the full Eq.(\ref{eq:12})
one expects that the excited states can be disentangled into
almost degenerate singlets and triplets, which in turn can be
interpreted as an excitation of the pseudo-scalar pion,
or the ground state of a vector meson, respectively.
In any case, the first excited state of the simple model correlates
very well with the mass of the $\rho^+$ and the other vector mesons,
see below.

\section{On the question of confinement}

The question of confinement is closely related to the potential
$V(r)$, but the relation is subtle.
One of the great advantages of light-cone quantization is the
additivity of the free part and the interaction.
Because of the dimensions of invariant mass squares the relation
of potential and kinetic energy is somewhat hidden.
In analogy to a classical Hamiltonian, the invariant mass-squared 
operator $M^2$ in Eq.(\ref{eq:20})
is interpreted as the square of a position and momentum dependent mass
\begin{eqnarray}
    M(r,p) = m_s\sqrt{1 + \frac{p ^{\,2}}{m_s m_r} + \frac{2}{m_s}V(r)}  
.\end{eqnarray}
For $V(r)\ll {m_s}$ and $p ^{\,2}\ll {m_s m_r}$
it could be expanded as

\begin{tabular}{llclclc} 
 $\displaystyle M(r,p)$  
 &$\displaystyle \simeq$ 
 &$\displaystyle m_s$
 &+
 &$\displaystyle \frac{p ^{\,2}}{2m_r}$
 &+
 &$\displaystyle V(r)$,
\\
 &$\displaystyle \simeq$
 &rest mass 
 &+
 &kinetic energy
 &+
 &potential energy,
\end{tabular}

\noindent
These conditions, however, can not be satisfied everywhere. 
I therefore propose to introduce a
{\em relativistic potential energy} by $W(r)\equiv M(r,p=0)$.
This function is plotted in Fig.~\ref{fig:6.2}.
It vanishes at $r_0=0.5194$~fm and the classical turning
point is $r_t=0.5300$~fm, in agreement with Eq.(\ref{eq:r}). 
Since the sum of the quark masses is 812~MeV,
the ionization threshold occurs at 542~MeV: 
In order to liberate the quarks in the pion,
one has to invest more than three times the rest energy of a pion. 

\section{The meson masses}

\begin{table} 
\begin{center}
\begin{tabular}{@{}c||rrrrr||rrrrr@{}} 
     \rule[-1em]{0mm}{1em} 
     & $\overline u$ & $\overline d$ 
     & $\overline s$ & $\overline c$ & $\overline b$ 
     & $\overline u$ & $\overline d$ 
     & $\overline s$ & $\overline c$ & $\overline b$  \\ \hline 
 $u$ & \rule[0.5em]{0mm}{1em}     
            & 768  & 892  & 2007 & 5325 
     &      & 768  & 871  & 2030 & 5418 \\
 $d$ & 140  &      & 896  & 2010 & 5325 
     & 140  &      & 871  & 2030 & 5418 \\
 $s$ & 494  & 498  &      & 2110 &  --- 
     & 494  & 494  &      & 2124 & 5510 \\
 $c$ & 1865 & 1869 & 1969 &      &  --- 
     & 1865 & 1865 & 1929 &      & 6580 \\
 $b$ & 5278 & 5279 & 5375 &  --- &      
     & 5279 & 5279 & 5338 & 6114 &          
\end{tabular}
\caption{\label{tab:2.1} 
   {\em Left:} Experimental masses of the flavor-off-diagonal 
   physical mesons in MeV.
   Scalar mesons are given in the lower, 
   vector mesons in the upper triangle.~-- {\em Right:} 
   The calculated mass eigenstates of QCD in MeV. 
   Singlet-1s states are given in the lower,
   singlet-2s states in the upper triangle.
}\end{center}
\end{table}

The remaining parameters of the theory, the masses of the
strange, charm and bottom quark can then be determined
by reproducing the masses of the pseudo-scalar mesons 
$K,^-$ $D^0$ and $B,^-$ respectively. This gives
$\overline m_s = 508$, $\overline m_c =1666$ 
and $\overline m_b =5054\mbox{ MeV}$.
The remaining off-diagonal pseudo-scalar mesons are then 
calculated straightforwardly and compiled in Table~\ref{tab:2.1}. 
In view of the simplicity of the model,
the agreement with the empirical values is remarkable, indeed, 
and at least as good as for any other model. 
The strong correlation of the first excited state with the 
vector mesons was mentioned.

\section{Summary and some conclusions} 

The very difficult problem of solving the many-body
Hamiltonian for QCD is replaced here by the problem of solving
an effective QCD-inspired Hamiltonian.
It is obtained from the QCD-Lagrangian in the light-cone gauge
by the method of iterated resolvents or the Hamiltonian flow equations
in the lowest non-trivial approximation.
In order to avoid the numerical complexities of the full solution
in light-cone coordinates, this effective interaction
was stripped-off of almost all ingredients except 
the Coulomb and the hyperfine interaction, which are
analyzed in a non-relativistic treatment with only 3-momenta. 
It is argued why the hyperfine interaction in the singlet channel
needs to be regulated like the usual divergencies in gauge field theory.

It was courageous to apply such a simple model to the
mystery particle of QCD, to the pion, but it turns out that
both the pion's mass and  size are obtained
without vacuum condensates of any kind, just
by adjusting the canonical parameters of a gauge field theory,
the coupling constant and the quark masses.
The necessary regularization constant determines itself by the theory, 
in line with the current interpretation of the renormalization group.
For the pion, as well as for the other mesons, one gets thus 
for the first time a QCD-inspired light-cone 
wave function by which one can compute the form factor
by the available exact formulas, 
as well as the higher Fock space amplitudes.

The present approach exposes screening rather than confinement. 
The removal energy of a quark from a meson is higher than the pion's mass. 
A bare quark is thus unlikely to be observed.
One wonders whether this is not much more a physical picture
than the strict linear confinement obtained by lattice gauge theory,
the only available alternative approach for a non-perturbative description.
But like the light-cone approach, the latter is burdened with its own
pecularities among them the far extrapolation down to the pion.

In fact the present model runs short in several aspects.
That the regulator function should apply to the whole kernel,
and not only to one piece of the interaction, is a technical detail, 
which can be easily corrected in future work.
More importent is whether the two-gluon annihilation
interaction, which was omitted here on perpose,
can account for the flavor diagonal mesons and the so important 
aspects of isospin.

Astounding is also that the present model for QCD differs from QED 
only by the color factor $\frac{4}{3}$ in the coupling constant. 
Can it be true that the essential difference between these two theories 
with their so distinctly different phenomenology is manifest only 
as large and small coupling?

Depite all of that 
I conclude that the light-cone approach to hadronic physics may now 
have a fair chance to compute the hadronic spectra and to predict 
the hadronic wave functions in line with a covariant theory.


\end{document}